\documentclass[12pt]{article}
\usepackage{psfig}
\usepackage{verbatim}
\usepackage{color}
\usepackage{rotate}
\usepackage{graphics}
\usepackage{epsf}
\usepackage{epsfig}
\usepackage{graphicx}
\usepackage{latexsym}
\usepackage{amscd}
\usepackage{amssymb}

\begin{document}

\begin{center}
{\Large
\bf
Non-abelian dynamics in first-order cosmological phase transitions}
\end{center}
\vspace{.5cm}

\begin{center}

 {\large Mikkel B. Johnson$^a$, L. S. Kisslinger$^b$, E. M. Henley$^c$, W-Y. P. Hwang$^d$, and T. Stevens$^e$}\\

\bigskip

{\sl $^a$Physics Division, Los Alamos National Laboratory, Los Alamos, NM 87545, USA}\\
                                                                                                                                                          
{\sl $^b$Department of Physics,  Carnegie Mellon University, Pittsburgh, PA 15213, USA}\\

{\sl $^c$Department of Physics, University of Washington, Seattle, WA 98195, USA}\\

{\sl $^d$Department of Physics, National Taiwan University, Taipei, Taiwan 106 R.O.C}\\

{\sl $^e$Department of Physics, New Mexico State University, Las Cruces, NM 88003, USA}\\

\end{center}

\vspace{1cm}

\begin{abstract} \noindent Bubble collisions in cosmological phase transitions are explored, taking
the non-abelian character of the gauge fields into account.  Both the QCD and electroweak
phase transitions are considered.  Numerical solutions of the field equations in several
limits are presented.
\medskip


\end{abstract}

\section{Introduction}	

The investigations reported in this talk have been motivated by an interest in studying cosmological phase
transitions quantitatively, taking the non-abelian character of the gauge fields into account.  Ultimately, 
we hope to identify observable consequences of cosmological phase transitions.

First-order phase transitions proceed by nucleation of bubbles of the broken phase in the background
of the symmetric phase.  Bubble collisions are of special interest, as they may lead to observable effects 
such as correlations in the cosmic microwave background (CMB)\cite{kisslinger} or as seeds of galactic 
and extra-galactic magnetic
fields\cite{kibble}.  The quantum chromodynamic (QCD) and the electroweak (EW) phase transitions
 are both candidates of interest in these respects.
The Lagrangian driving both the QCD and the EW phase transitions are essentially known
and make it possible to approach the physics of the phase transitions from first principles.
However, a difficulty to making reliable predictions is that the fundamental guage fields in both these
instances are non-abelian:  the gluon field in QCD and the $W$ and $Z$ fields in the EW case.
The quantitative role of non-abelian fields in cosmological phase transitions is poorly known and 
difficult to calculate due to the nonlinearities arising from the 
non-abelian character of the gauge fields.  

\section{A Toy Model}
The results reported in this talk are preliminary and correspond to equations of motion that follow from toy model Lagrangian,
\begin{eqnarray}
\label{L}
  {\cal L} & = & {\cal L}^{(1)} + {\cal L}^{(2)}
\\ \nonumber
         {\cal L}^{(1)} & = & -\frac{1}{4}W^i_{\mu\nu}W^{i\mu\nu}
  -\frac{1}{4} B_{\mu\nu}B^{\mu\nu} \\ \nonumber
 {\cal L}^{(2)} & = & |(i\partial_{\mu} -\gamma\frac{g}{2} \tau \cdot W_\mu
 - \frac{g'}{2}B_\mu)\Phi|^2  -V(\Phi) \, ,
\end{eqnarray}
with
\begin{eqnarray}
\label{phi}
         \Phi(x) & = & \left( \begin{array}{clcr} 0 \\
                                \phi(x)
         \end{array} \right) 
           =  \left( \begin{array}{clcr} 0 \\
                                f(x)e^{i\Theta(x)}
         \end{array} \right)~,
\end{eqnarray}
\begin{eqnarray}
\label{wmunu}
  W^a_{\mu\nu} & = & \partial_\mu W^a_\nu - \partial_\nu W^a_\mu
 - g \epsilon_{abc} W^b_\mu W^c_\nu\\ \nonumber
 B_{\mu\nu} & = & \partial_\mu B_\nu -  \partial_\nu B_\mu \, .
\end{eqnarray}
The potential $V(\Phi)$ in Eq.~(\ref{L}) leads to spontaneous symmetry breaking that will drive
the first-order EW phase transition of interest in this work.  The detailed form of $V(\Phi )$ depends upon the theory,
but one possible form is given by
\begin{eqnarray}
\label{V}
    V(\Phi ) \equiv V(f)= & -\mu^2 |\Phi|^2 + \lambda |\Phi|^4 \, .
\end{eqnarray}

This Lagrangian consists of two coupled sectors.  One of these is abelian, consisting of the scalar boson $\Phi$ 
and the vector field $B^a_\mu$, which are the familiar scalar Higgs and the vector boson of the weak 
hypercharge current, 
respectively, of the Weinberg-Salam model\cite{halzen}.  A first-order EW phase transition requires that the Lagrangian 
be considered in the framework of the minimal supersymmetric model (MSSM) extension of the 
Weinberg-Salam model, and one may eventually need to explicitly incorporate the {\it stop} field for consistency,
as discussed in more detail in Ref.~\cite{kisslinger}.  The other sector is non-abelian 
and consists
of the vector field $W^a_\mu$.  We will interpret this field as the gluon field of $SU(2)_{color}$ 
in our study of the QCD phase transition in Sect.~3 and as the three vector gauge bosons of the Weinberg-Salam
model in our study of the of the EW phase transition in Sects.~4 and 5.  This interpretation is possible, 
in the spirit of our toy model, since both fields satisfy the same field equations in the absence of coupling 
($\gamma =0$) between the sectors.  Fermions do not appear in the toy Lagrangian, which reflects our focus so far on 
issues of non-abelian dynamics that arise in their absence.  

Equations of motion are obtained by minimizing the action,
\begin{eqnarray}
\label{action}
 \delta [\int d^4x( {\cal L}^{(1)}+{\cal L}^{(2)})] & = & 0~.
\end{eqnarray} 
The result of doing this yields two ``$W$-equations":
\begin{eqnarray}
\label{eomw3}
\partial^2 W^3_\nu-\partial_\mu \partial_\nu  W^3_\mu
 -g \epsilon_{3jk}[ W^k_\nu \partial_\mu  W^j_\mu
 +2 W^j_\mu \partial_\mu  W^k_\nu
 - W^j_\mu \partial_\nu  W^k_\mu]) \\ \nonumber
  +g^2 \epsilon_{3jk} W^j_\mu \epsilon^{klm} W^l_\mu W^m_\nu
   -f(x)^2\gamma g\psi_\nu(x) & = & 0 \, ,~~i=3,
\end{eqnarray}
\begin{eqnarray}
\label{eomw12}
\partial^2 W^i_\nu-\partial_\mu \partial_\nu  W^i_\mu
 -g \epsilon_{ijk}[ W^k_\nu \partial_\mu  W^j_\mu
 +2 W^j_\mu \partial_\mu  W^k_\nu
 - W^j_\mu \partial_\nu  W^k_\mu]) \\ \nonumber
  +g^2 \epsilon_{ijk} W^j_\mu \epsilon^{klm} W^l_\mu W^m_\nu
   +\frac{\gamma}{2}f(x)^2 g^2 W^i_\nu & = & 0 \, ,~~i=(1,2),
\end{eqnarray}
a ``$\Theta$-equation",
\begin{equation}
\label{eomth}
\partial^\alpha f(x)^2\psi_\alpha (x)=0,
\end{equation}
a ``$B$-equation",
\begin{equation}
\label{eomb}
\partial^2B_\nu-\partial^\mu\partial_\nu B_\mu+f(x)^2g'\psi_\nu (x) =0,
\end{equation}
and an ``$f$-equation",
\begin{equation}
\label{eomf}
\partial^2f(x)-\frac{\gamma g^2}{4}f(x)[(W^1_\nu)^2+(W^2_\nu)^2]-f(x)\psi_\alpha\psi^\alpha+f(x)\frac{\partial V}
{\partial f^2}=0~.
\end{equation}
The quantity $\psi_\alpha$ is defined as
\begin{equation}
\label{eompsi}
\psi_\alpha (x)\equiv \partial_\alpha\Theta+\frac{g'}{2}B_\alpha-\frac{\gamma g}{2}W^3_\alpha~.
\end{equation}

\section{Bubble Collisions in QCD} 
As discussed in Sect.~2, the equation of motion for the gluon field $A^a_\mu$ in $SU(2)_{color}$ is the same as that
of $W^a_\mu$ above in the absence of coupling between the abelian and non-abelian sectors, {\it i.e} $\gamma =0$.
In the gauge $\partial^\mu A^a_\mu =0$, this is
\begin{eqnarray}
\label{EOM_A}
\partial_\mu\partial^\mu A^a_\nu 
+ g\epsilon^{abc}(2 A^b_\mu\partial^\mu A^c_\nu                         
- A^b_\mu \partial_\nu A^{\mu c})                                       
+ g^2\epsilon^{abc}\epsilon^{cef}A^b_\mu A^{\mu e}A^f_\nu = 0.          
\end{eqnarray}
This equation has a well-known non-perturbative solution, namely the BPST instanton\cite{bpst},
\begin{eqnarray}
\label{A_inst1}
A^{a }_\mu(x) &=& \frac{2}{g}\frac{\eta_{a\mu\nu}x^\nu}{x^2+\rho^2}~.
\end{eqnarray}
Using an instanton-like ansatz in 1+1 dimensions, 
\begin{eqnarray}
\label{A_inst2}
A^{a }_\mu(x,t) &=& \frac{2}{g}\eta_{a\mu\nu}x^\nu F(x,t)~, 
\end{eqnarray}
bubble collisions were found in Ref.~\cite{jkc} to evolve as solutions of the equation 
\begin{eqnarray}
\label{eominst}                                
   \partial^2 F & = & -\frac{2}{x}\partial_x F -12 F^2 +8 (x^2-t^2)F^3~
\end{eqnarray}  
in Minkowski space, assuming periodic boundary conditions in time with
\begin{eqnarray}
\label{initial}
  F(x,0)& = & \frac{1}{(x-3)^2 + \rho^2} + \frac{1}{(x+3)^2 + \rho^2}\\
\nonumber
  \partial_t F(x,0) & = & 0~.
\end{eqnarray}

\begin{figure}
\centerline{\psfig{figure=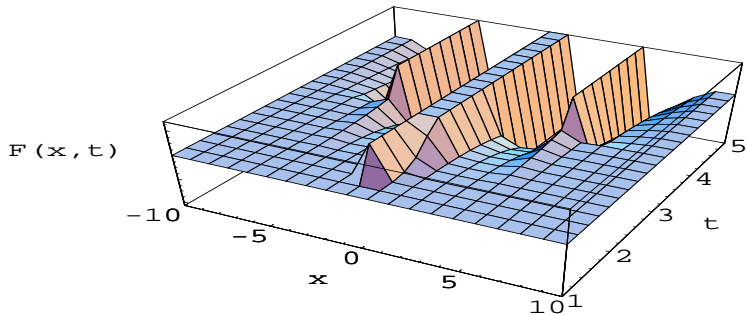,height=5cm,width=10cm}}
\caption{F(x,t) for the two bubbleswith instanton-like form}
\label{Fig.1}
\end{figure}

The numerical solution to Eq.~(\ref{eominst}) is given in Fig. \ref{Fig.1}. Note the
development of a gluonic wall at the x=0 collision region. 
The wall grows rapidly beginning at time $t=1.1$, and
due to the singularities in the solution the accuracy of the calculations
for t $>$ 1.0 is limited.  A possible connection to CMB correlations is discussed in Ref.~\cite{kisslinger}.

\section{Bubble Collisions in the Abelian Higgs Model}
The abelian Higgs model has been of interest as a prototype for the generation of magnetic fields in
the early universe in collisions of bubbles during a first-order EW phase 
transition\cite{kibble,AE,CST}.  The Lagrangian of the abelian Higgs model describes a complex scalar field coupled to the 
electromagnetic (em) field $ A^{em}_\mu$.  It is defined by the Lagrangian Eq.~(\ref{L}) in the the abelian sector 
identifying $ A^{em}_\mu$ with the field $B_\mu$ and the electric charge $e$ with the coupling 
parameter $g'$ as $e=g'/2$.  Equations of motion for the Higgs and magnetic field may be read off the results in 
Eqs.~(\ref{eomth}, \ref{eomb},\ref{eomf}) for $\gamma =0$, {\it i.e.},
\begin{equation}
\partial^\alpha f(x)^2\psi_\alpha (x)=0~,
\end{equation}
\begin{equation}
\partial^2B_\nu-\partial^\mu\partial_\nu B_\mu+2ef(x)^2\psi_\nu (x) =0,
\end{equation}
and 
\begin{equation}
\partial^2f(x)-f(x)\psi_\alpha\psi^\alpha+f(x)\frac{\partial V}
{\partial f^2}=0~.
\end{equation}
The quantity $\psi_\alpha$ is now
\begin{equation}
\psi_\alpha (x)= \partial_\alpha\Theta+eA^{em}_\alpha~.
\end{equation}

In this case, the phase transition is driven by the dynamics of the Higgs field along the lines studied by 
Coleman\cite{coleman}.  Coleman studied the case of a real scalar field ($\Theta =0$), in which case the
equation of motion for the scalar field becomes
\begin{equation}
\label{fcoleman}
\partial^2f(x)+f(x)\frac{\partial V}
{\partial f^2}=0~.
\end{equation}
The potential $V(f)$ given in Eq.~(\ref{V}) describes the dependence of the energy of vacuum on the scalar 
field.  It has two minima, corresponding to ``true" and ``false" vacua.  In Coleman's model, a symmetry breaking 
term is added to $V(f)$ to give the true vacuum a slightly lower energy.  

One imagines that the system (here, the universe) begins in the false vacuum, and then, as time evolves, a transition 
is made to the true vacuum.  The phase transition proceeds as bubbles of the true vacuum nucleate in the false vacuum.  
Nucleated bubbles are tunneling (instanton) solutions of the $f$-equation in Euclidean space\cite{coleman}.  
Once nucleated, bubbles grow and collide as Minkowski space solutions of the $f$-equation.  

In their analysis of the abelian Higgs model, Kibble and Vilenkin\cite{kibble} suggested one way in which magnetic fields 
might be generated as bubbles collide. They considered the regime of gentle collisions, where $f(x)$ remains 
constant, or nearly constant, in the region of overlap of the colliding bubbles.  They were able to gain insight into the 
generation of magnetic fields in this case by making an expansion about point $f(x)=f_0$, which we shall refer to as
the Kibble-Vilenkin point.  For the case of gentle collisions, we can assume the following expansion,
\begin{eqnarray}
\label{kv1}
f(x)&=&f_0+a\delta f(x)\\ 
\psi_\alpha(x)&=&a\psi^{(1)}+a^2\psi^{(2)}_\alpha +~...~, \nonumber
\end{eqnarray}
where $a$ is the magnitude of $f_0-f(x)$ and is small by assumption.  Substituting the expansion into the equations 
of motion and requiring that the equations be satisfied at each order in the expansion parameter $a$, the relevant 
$\Theta$ and $B$ equations give, to leading order in $a$, the following results
\begin{eqnarray}
\label{kv2}
(\partial^2+2e^2f_0^2)\psi^{(1)}_\alpha=0\\
\label{kv3}
\partial^\alpha\psi^{(1)}_\alpha=0~,
\end{eqnarray}
where now
\begin{eqnarray}
\label{kv4}
\psi^{(1)}_\alpha=\partial_\alpha\Theta+eA^{em(1)}_\alpha~.
\end{eqnarray}
These are essentially the equations of Kibble and Vilenkin\cite{kibble}, who demonstrated from them that magnetic fields 
encircle the overlap region of the colliding bubbles when the phase of the Higgs fields is initialy different within each 
bubble.  Their analysis has been elaborated upon by Copeland, Saffin, and T\"ornkvist\cite{CST}, who presented solutions in 
a convenient, closed form.

\begin{figure}
\centerline{\psfig{figure=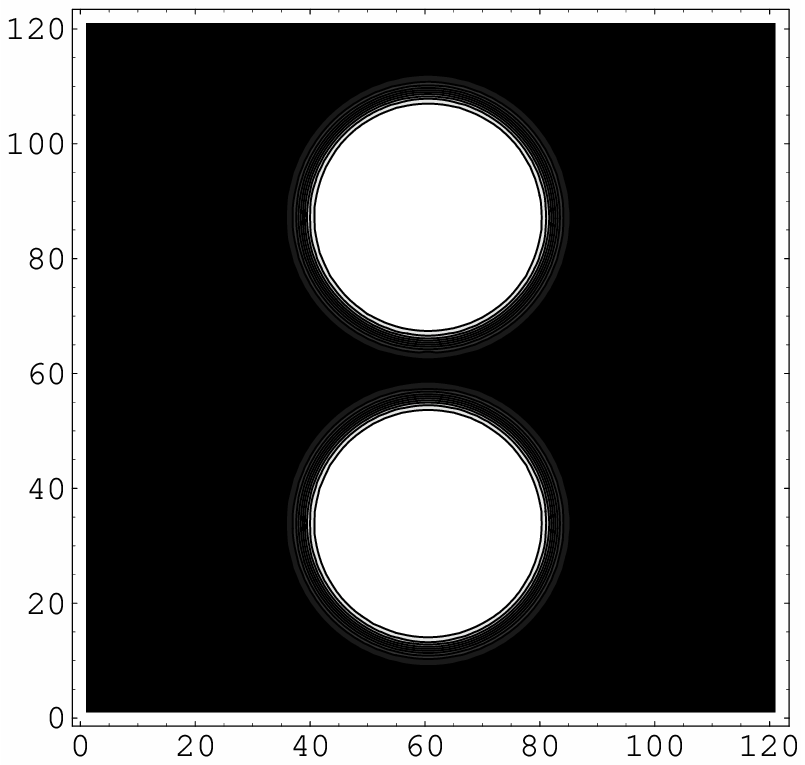,height=4cm,width=4cm}~~~~~~~\psfig{figure=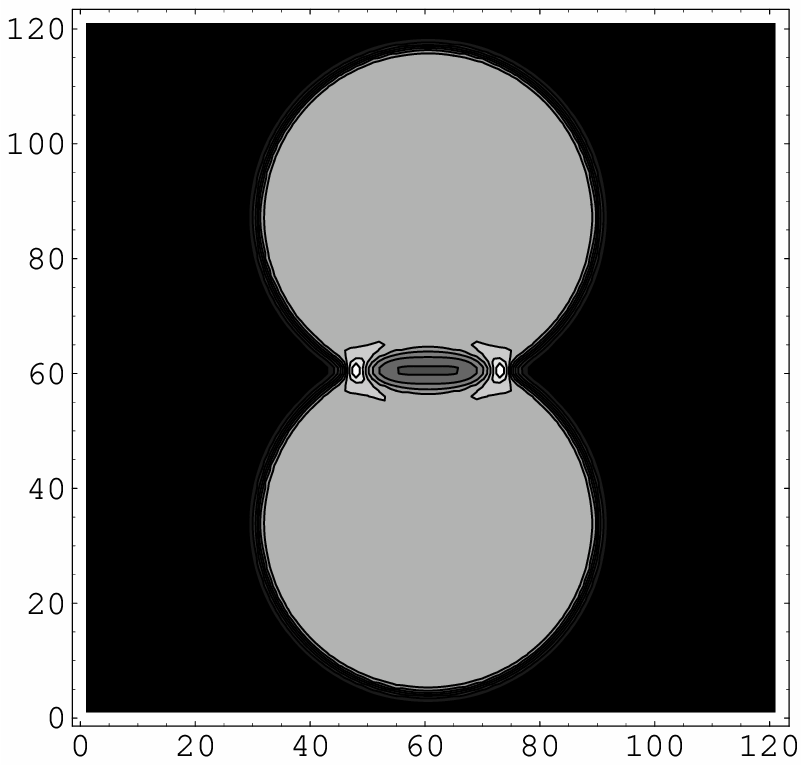,height=4cm,width=4cm}}
\caption{Two bubbles colliding in the Coleman model as explained in the text}
\label{Fig.2}
\end{figure}

For violent collisions, where $f(x)$ changes substantially in the collision, the character of the problem requires 
numerical integration of the full coupled PDE.  We close this section by showing numerical solutions of Eq.~(\ref{fcoleman}) 
using an algorithm for solving the PDE in 2 + 1 dimensions.  The initial condition is shown in the left-hand panel
of Fig. \ref{Fig.2}, and the solution of the equations of motion just after the collision begins is shown in 
the right-hand panel.  The solutions have been followed sufficiently far in time 
to convince us that the algorithm is stable, and it is easily generalized to include multiple coupled fields.  

\section{Bubble Collisions in the EW Phase Transition}
One of our main interests is to understand the generation of the em field during the EW phase transition,
specifically when non-abelian gauge fields play a role.  In this case, $\gamma =1$, and the $W$, $B$, 
and $\phi$ (Higgs) fields are fully coupled.  The physical $Z$ and $A^{em}$ fields are 
determined in terms of these fields as
\begin{eqnarray}
\label{Aem}
A^{em}_\mu = \frac{1}{(g^2+g'^2)^{1/2}}(g'W^3_\mu+gB_\mu)\\
\label{Z}
Z_\mu= \frac{1}{(g^2+g'^2)^{1/2}}(gW^3_\mu-gB_\mu).
\end{eqnarray}
We consider here only the case of 
gentle collisions, extending the analysis given by Kibble and Vilenkin for the abelian Higgs model.  In the
future we will examine the case of more violent collisions solving the equations of motion numerically.

As in the abelian Higgs model, the MSSM equations simplify upon expansion about the Kibble-Vilenkin point,
\begin{eqnarray}
f(x)=f_0+a\delta f(x). 
\end{eqnarray}
The fact that $\psi$ and $W^d$ (for $d=(1,2)$) enter quadratically in the $f$-equation places two important constraints on 
these quantities:  (1) $\psi$ and $W^d$ (for $d=(1,2)$) must have an expansion in odd powers of $a^{1/2}$, if we
require the square of these quantities be analytic in $a$; and, (2) expanding this equation to leading order in $a^{1/2}$,
we find that the terms $\psi^{(0)}$, $w^{(0)1}$, and $w^{(0)2}$ must vanish.  This is most easily seen in the Euclidean 
metric, from the fact that the square of each enters with the same sign.  However, the same must be true in the Minkowski 
metric as well by analytic continuation.  In view of these considerations, $\psi_\alpha$ and 
$W^d_\alpha$ for $d=(1,2)$ have the following expansion
\begin{eqnarray}
\psi_\alpha(x)=a^{1/2}\psi^{(1)}+a^{3/2}\psi^{(3)}_\alpha +~...\\
W^d_\alpha=a^{1/2}w^{(1)d}_\alpha+a^{3/2}w^{(3)d}_\alpha +~...~.
\end{eqnarray}
It is natural that an expansion in the same parameter $a^{1/2}$ remains appropriate for $d=3$.  However,
there is no requirement that the leading term vanish, so we take
\begin{eqnarray}
W^3_\alpha=w^{(0)3}_\alpha+a^{1/2}w^{(1)3}_\alpha+a^{3/2}w^{(3)3}_\alpha +~...~.
\end{eqnarray}

The $B$-, $\Theta$-, and $W$-equations then give, to first order in $a^{1/2}$, 
\begin{eqnarray}
\label{eompsi1}
[\partial^2+\frac{f_0^2}{2}(g^2+g'^2)]\psi^{(1)}_\alpha=0\\
\label{eompsi2}
\partial^\alpha\psi^{(1)}_\alpha=0
\end{eqnarray}
where now
\begin{eqnarray}
\label{psi3}
\psi^{(1)}_\alpha(x)=\partial_\alpha\Theta-\frac{(g^2+g'^2)^{1/2}}{2}Z^{(1)}_\alpha~.
\end{eqnarray}
Comparing these equations to those in the abelian Higgs model, Eqs.(\ref{kv2},\ref{kv3},\ref{kv4}), we see that the 
$Z$ field here plays the same role as the em field did in that case.  Specifically, in this case, the phase 
difference of the Higgs field now determines the gauge field $Z_\mu$ of Eq.(\ref{Z}) within the bubble 
overlap region.  Thus, for gentle collisions,  the mathematical problem in leading order 
is no more complicated that it was in the abelian case.  

We may obtain $A^{em}$ from Maxwell's equation, formed by taking the linear combination of the $W^{(3)}$- 
and $B$-equations, Eqs.(\ref{eomw3}) and (\ref{eomb}), suggested by Eq~(\ref{Aem}).  We find that the first non-vanishing
contribution to the em current occurs at order $a^{3/2}$ and depends cubically upon the non-abelian 
fields $w^{(1)d}_\alpha$ calculated at order $a^{1/2}$.

Equations for $w^{(1)d}_\alpha$ may be obtained by expanding the $B$- and $W$-equations through order $a^{1/2}$.
We find for $d=3$, 
\begin{eqnarray}
\label{eomu30}
\partial^2u^{(0)3}_\nu-\partial_\nu\partial\cdot u^{(0)3}=0
\end{eqnarray}
and
\begin{eqnarray}
\label{eomu31}
\partial^2u^{(1)3}_\nu-\partial_\nu\partial\cdot u^{(1)3}=\frac{f_0^2g^2}{2}\psi^{(1)}_\nu~,
\end{eqnarray}
where we have expressed the equations in terms of $u_\nu$ defined as
\begin{eqnarray}
u_\nu(x)=\frac{g}{2}w_\nu(x)~.
\end{eqnarray}
Note that $u^{(1)3}_\nu(x)$ follows from Eq.~(\ref{eomu31}) once the driving term 
$\psi^{(1)}(x)$ has been independently determined from Eq.~(\ref{psi3}) and the solution of 
Eqs.~(\ref{eompsi1},\ref{eompsi2}).  For $d=$ 1 or 2 (corresponding to $d^{\prime}=$ 2 or 1, respectively), we obtain
the pair of equations
\begin{eqnarray}
\label{eomu12}
\partial^2u^{(1)d}_\nu&-&\partial_\nu\partial\cdot u^{(1)d}-2[\partial^\mu(u^{(0)3}_\nu u^{(1)d^{\prime}}_\mu-
u^{(1)d^{\prime}}_\nu u^{(0)3}_\mu)\\ \nonumber
&+&(u^{(1)d^{\prime}}_\mu\partial^\mu u^{(0)3}_\nu -u^{(0)3}_\mu\partial^\mu u^{(1)d^{\prime}}_\nu)-
(u^{(1)d^{\prime}}_\mu\partial_\nu u^{(0)\mu 3} -u^{(0)3}_\mu\partial_\nu u^{(1)\mu d^{\prime}})]\\ \nonumber
&-&4[(u^{(0)3})^2u^{(1)d}_\nu-u^{(0)3}\cdot u^{(1)d}u^{(0)3}_\nu]+\frac{f_0^2g^2}{2}u^{(1)d}_\nu=0~.
\end{eqnarray}
Note that Eq.~(\ref{eomu12}) is a linear equation for the non-abelian fields $u^{(1)d}$.  The non-abelian field 
$u^{(0)3}$ enters nonlinearly, but it is determined from a separate, uncoupled equation, 
Eq.~(\ref{eomu30}).  Thus, for sufficiently gentle collisions, {\it all} relevant equations are linear.  This means,
among other things, that one can avoid introducing a nonperturbative instanton/sphaleron ansatz as in Eq.~(\ref{A_inst2}),
which would lead to equations that do not transform properly under a lorentz transformation.

\section{Summary and Conclusions}
Methods suitable for exploring the role of non-abelian dynamics in QCD and EW phase transitions have been
explored.  The QCD phase transition is essentially non-abelian, and our numerical investigations have been 
carried out using a BPST instanton-form solution to simplify the treatment of the non-linear dynamics.  In the case of the 
EW phase transition, we found that for 
gentle bubble collisions the non-abelian fields may be obtained by solving {\it linear} equations, so that a
nonperturbative sphaleron ansatz is not needed to account for the nonperturbative dynamics.  Investigations 
are continuing for the case of gentle collisions along the lines outlined here. Our preliminary numerical work 
for the Coleman model suggests that pursuing numerical solutions to the PDE is a promising approach for
determining the consequences of more violent collisions.


\begin{thebibliography}{0}
\bibitem{kisslinger}  See the contribution by Leonard Kisslinger at this conference.
\bibitem{kibble}  T. W. B. Kibble and Alexander Vilenkin, {\it Phys. Rev. D} {\bf 52}, 679 (1995).
\bibitem{halzen}  F. Halzen and A. D. Martin, {\it Quarks and  Leptons:  An Introductory Course in Modern Particle Physics}
(John Wiley and Sons, 1984).
\bibitem{bpst}  A. A. Belavin, A. M. Polyakov, A. S. Schwartz, and Yu. S. Tyupkin, {\it Phys. Lett} {\bf B59}, 85 (1975).
\bibitem{jkc}  M. B. Johnson, H.-M. Choi, and L. S. Kisslinger, {\it Nucl. Phys. A} {\bf 729}, 729 (2003).
\bibitem{AE} J. Ahonen and K. Enqvist, {\it Phys. Rev. D} {\bf 57}, 664 (1997).
\bibitem{coleman}  S. Coleman, {\it Aspects of Symmetry} (Cambridge, 1985), Ch. 7.
\bibitem{CST}  E. J. Copeland, P. M. Saffin, and O. T\"ornkvist, {\it Phys. Rev. D} {\bf 61}, 105005 (2000).
\end{thebibliography}
\end{document}